\input{aipcheck}
\documentclass[final]{aipproc}

\layoutstyle{6x9}
\usepackage{amsmath}
\usepackage{grffile}
\newcommand{\vmean}[1][]{\langle v#1 \rangle}
\begin{document}
\title{Creating Non-Maxwellian Velocity Distributions in Ultracold Plasmas}
\author{J. Castro}{
  address={Department of Physics and Astronomy and Rice Quantum Institute, Rice University, Houston, Texas 77005, USA}
}
\author{G. Bannasch}{
  address={Max Planck Institute for the Physics of Complex Systems, 01187 Dresden, Germany}
}
\author{P. McQuillen}{
  address={Department of Physics and Astronomy and Rice Quantum Institute, Rice University, Houston, Texas 77005, USA}
}
\author{T. Pohl}{
  address={Max Planck Institute for the Physics of Complex Systems, 01187 Dresden, Germany}
}
\author{T. C. Killian}{
  address={Department of Physics and Astronomy and Rice Quantum Institute, Rice University, Houston, Texas 77005, USA}
}
\begin{abstract}
We present   techniques to perturb, measure and model the ion velocity distribution in an ultracold neutral plasma produced by photoionization of strontium atoms.
By optical pumping with circularly polarized light we promote ions with certain velocities to a different spin ground state, and probe the resulting perturbed velocity distribution through laser-induced fluorescence spectroscopy. We discuss various approaches to extract the velocity distribution from our measured spectra, and assess their quality through comparisons with molecular dynamic simulations.
\end{abstract}
\maketitle
\section{Introduction}
The ability to modify and probe the velocity distribution of particles in plasmas is valuable for studying collective modes \cite{add09}, transport \cite{css95}, and
thermalization rates in plasmas \cite{jhb05}.   Here, we describe the combined application of optical pumping and laser-induced fluorescence (LIF) Doppler spectroscopy to create and characterize perturbed ion velocity distributions in ultracold plasmas.
Optical pumping \cite{hap72} between different hyperfine states provides a powerful method for tagging certain particles in equilibrium systems.  Exploiting the Doppler effect, the spin-state modification can be done in a velocity-dependent manner to modify the velocity distribution of a given spin state. On the other hand,  LIF spectroscopy is a well established tool for measuring velocity distributions \cite{abj00} and has been used to probe pure ion plasmas \cite{add09} or plasmas created with short pulse lasers \cite{mla10}.

Ultracold plasmas (UCPs), produced by photo-ionization of laser-cooled atoms \citep{killian1999unpCreation} or cold molecular beams \citep{grant2008supersonic} provide a well controlled laboratory to study various plasma physics phenomena, such as collective waves \cite{bes03,fzr06,mlt09,cmk10,lpr10,mes11,shu11b}, plasma
expansion into vacuum \cite{kkb00,roh02,roh03,ppr03,ppr04b,lgs07,tro10,grant2009nobeam}, correlation effects \cite{mur01,mck02,kun02,ppr04a,scg04,csl04,ppr05,cdd05,shu11}, recombination to form neutral atoms
\cite{klk01,gls07,fletcher2007tbr_temp,pohl2008rates,br08,bap11}, and plasma instabilities \cite{zfr08,ros11}.
Due to their very low temperatures UCPs also realize an interesting parameter regime \cite{kpp07,Killian07,kir10} in which the ionic plasma component can be strongly coupled. Consequently, the present work opens up new studies of non-equilibrium plasma dynamics in the strong coupling regime.

\section{Creation of Ultracold Neutral Plasmas}
\begin{figure}
\begin{centering}
\includegraphics[clip,width=2.7in,angle=270]{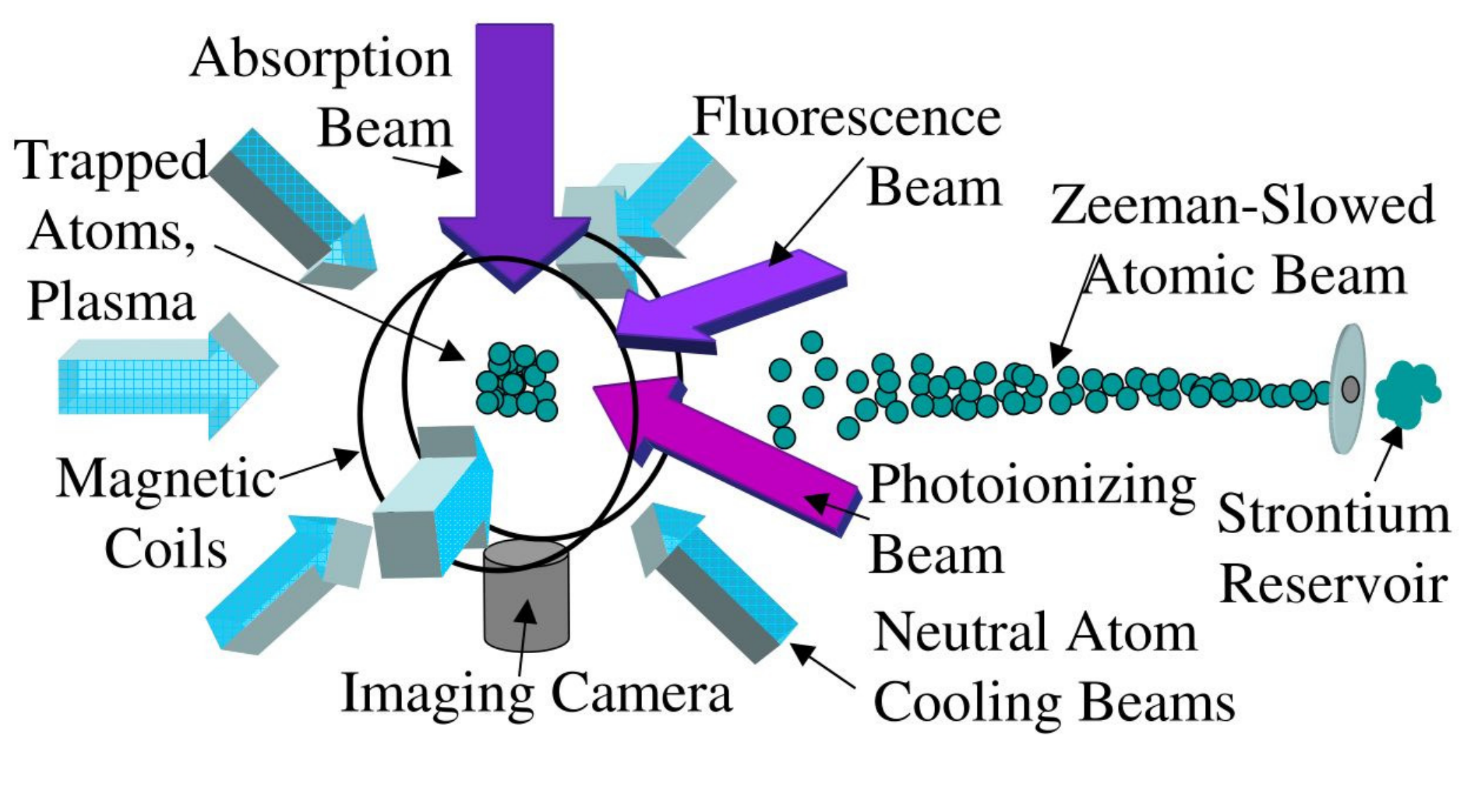}
\par\end{centering}
\caption{Schematics of the experimental setup. Neutral
strontium atoms from a heated reservoir are Zeeman-slowed before entering the trapping
region. The magneto-optical trap consists of a pair of anti-Helmholtz
magnetic coils and six laser beams. $^{1}\textrm{P}_{1}$
trapped atoms are then ionized by the photoionizing laser. The fluorescence
probe beam propagates in a direction that is perpendicular to the
imaging axis and CCD camera. The complementary absorption probe beam
passes through the plasma and falls on the camera. Adapted from \cite{scg04}.
\label{fig:apparatus}}
\end{figure}

We create an ultracold neutral plasma through photoionization of laser-cooled strontium
atoms in a magneto-optical trap (MOT) \cite{scg04,mvs99,nsl03}.
Figure \ref{fig:apparatus} shows our experimental setup.
The MOT operates on the dipole-allowed $^{1}\textrm{S}_{0}-^{1}\textrm{P}_{1}$ transition
of $^{88}$Sr at $461$ nm, with transition linewidth, $\gamma / (2\pi) = 30.5$ MHz \cite{nms05}.
The laser-cooled atom cloud is characterized by a temperature of $\sim10$ mK and has a
spherically symmetric Gaussian density distribution, $n({r})=n_{0}{\rm
exp}(-r^{2}/2\sigma^{2})$, with $\sigma\approx0.6$ mm and $n_{0}\approx6\times10^{10}$
cm$^{-3}$. The number of trapped atoms is typically $2\times10^{8}$.

\begin{figure}
\begin{centering}
\includegraphics[clip,height=2.4in]{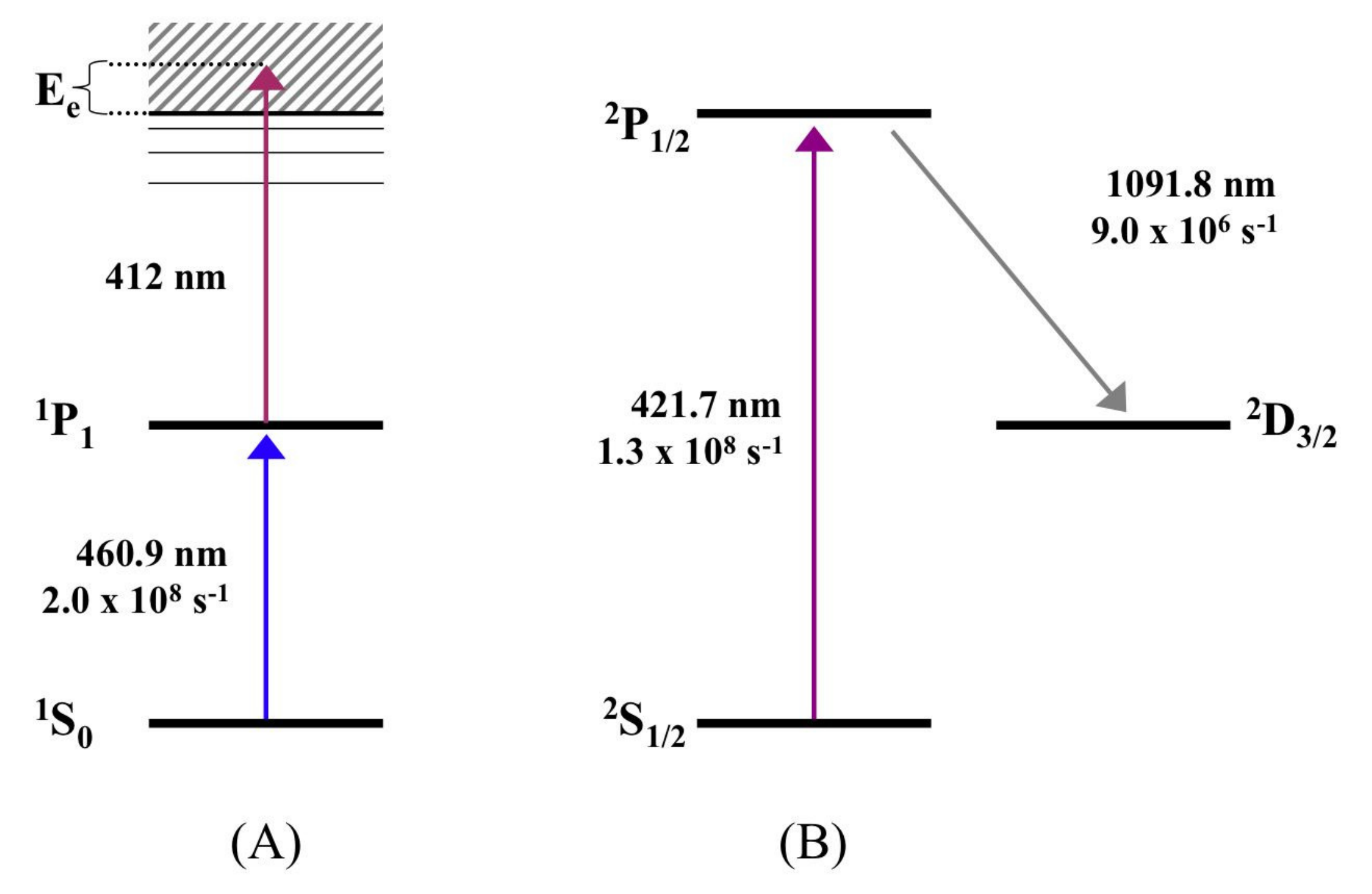}
\par\end{centering}
\caption{Relevant atomic and ionic energy levels of strontium,
with corresponding decay rates. (A) Neutral atoms are laser-cooled and trapped in
a MOT operating on the $^{1}\textrm{S}_{0}-{}^{1}\textrm{P}_{1}$
transition at 460.9 nm \cite{nsl03}.
Atoms are excited to the $^{1}\textrm{P}_{1}$ level by a pulse-amplified
laser and ionized by pulsed dye laser at $\sim412$ nm. (B) Imaging
of the ions is done on the $^{2}\textrm{S}_{1/2}-{}^{2}\textrm{P}_{1/2}$
transition at $421.7$ nm. $^{2}\textrm{P}_{1/2}$ ions decay to the
$^{2}\textrm{D}_{3/2}$ state 7\% of the time, after which they cease
to interact with the probe beam. The intensity and duration of the
$422$ nm light is sufficiently low to avoid optical pumping to the metastable
$^{2}\textrm{D}_{3/2}$ state.
\label{fig:energylevels}}
\end{figure}

The atoms are ionized via two-photon ionization by two temporally and spatially overlapping, retro-reflected $\sim$10 ns laser pulses: the first is obtained from a pulse-amplified laser beam tuned to the cooling transition of the atoms $(^{1}\textrm{S}_{0}-^{1}\textrm{P}_{1})$ at $461$ nm and the second one derives from a pulsed dye laser tuned just above the ionization continuum at $\sim412$ nm (Figure \ref{fig:energylevels} A).

In this way we ionize $\sim$30-70\% of the atoms. The plasma inherits its density distribution from the neutral atoms, resulting in peak electron and ion densities as high as $n_{0e}\approx n_{0i}\approx4.2\times10^{10}$ cm$^{-3}$. The remaining ground state atoms have no effect on the subsequent plasma dynamics, due to the short time scale of the experiment and the small neutral-ion collision cross-sections.

As a result of the much lighter mass of the electrons (compared to the heavy Sr ions),
most of the excess energy from the photoionizing beam is acquired by the electrons, while
the ions' kinetic energy remains similar to that of the neutral atoms in the MOT
\cite{kpp07}.  By tuning the wavelength of the pulsed dye laser, we vary the initial
electron kinetic energies typically between $1$ and $1000$ K. The ion kinetic energy
remains on the order of millikelvin, close to the kinetic energy of neutral atoms in the
MOT. However, disorder-induced heating \cite{mur01,csl04,ppr05} raises the temperature of the
cold ions to approximately 1 K on the timescale of the inverse ion plasma oscillation
frequency ($\sim 500$\,ns).

\section{Optical modification of velocity distributions}
\begin{figure}
  \begin{centering}
    \includegraphics[clip,width=3in]{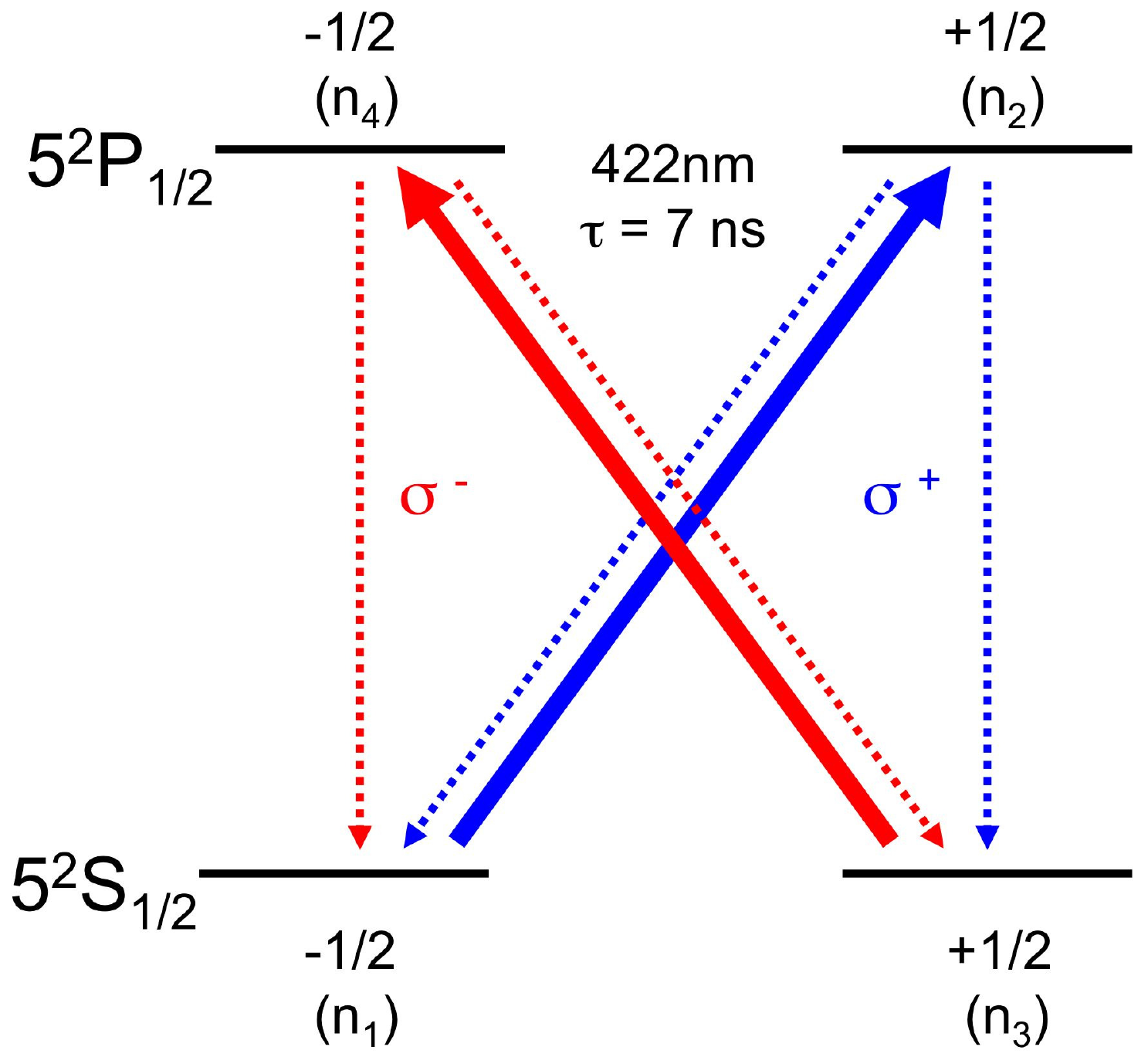}
  \par\end{centering}
  \caption{\label{fig:energy} Simplified energy level schematic for the \textit{Sr}$^{+}$,
    where dashed lines indicate decay paths while solid lines indicate
    excitation paths. The $m_j=-1/2$ $\left(m_j=+1/2\right)$ spin ground level is coupled
    to the higher energy $m_j=+1/2$ $\left(m_j=-1/2\right)$ via $\sigma^{+}$ $\left(\sigma^{-}\right)$
    circularly polarized laser beam, from where it can decay back to either
    the $m_j=-1/2$ or $m_j=+1/2$. $n_i$ labels the population of a particular spin state.}
\end{figure}

The internal structure of the strontium ions relevant to our experiment can be represented by the simplified four level scheme depicted in Figure \ref{fig:energy}.
Both the ground state ($5^2\textrm{S}_{1/2}$) and the excited state ($5^2\textrm{P}_{1/2}$) are doubly degenerate, where the total angular momentum projection can take on the
values $m_j = +1/2$ and $m_j = -1/2$. The two ground states of the $5^2\textrm{S}_{1/2}$
level will be denoted by $1$ $(m_j=-1/2)$ and $3$ $(m_j=+1/2)$, and the two excited states
of the $5^2\textrm{P}_{1/2}$ level will be denoted by $2$ $(m_j=+1/2)$ and $4$
($m_j=-1/2$).

The transitions between level 1 and 2 and between level 3 and 4 are driven by lasers with
circularly polarized light, $\sigma_{+}$ and $\sigma_{-}$ respectively, with frequencies
$\omega_{ij}$ and Rabi frequencies
\begin{equation}
  \Omega_{ij} = \frac{d_{ij} E_0}{\hbar},
  \label{rabi}
\end{equation}
where $E_0$ is the field amplitude of the laser. The corresponding frequency detuning for each transition is $\Delta_{ij} = \omega_{ij} - \omega$, where
$\hbar \omega = \Delta E$ is the energy difference between ground and excited state.

Ions in the excited states ($2$ and $4$) decay to the ground states ($1$ and $3$) with decay rates
\begin{equation}
  \Gamma_{ij} = \frac{8 \pi^2 d_{ij}^2}{3 \epsilon_0 \hbar \lambda^3}\\
  \label{gamma}
\end{equation}
where $d_{ij}$ is the dipol matrix element of the optical transition between level $i$ and
$j$, $\lambda$ is the wavelength of the transition between ground and excited state, $\epsilon_0$ is the electric constant and
$\hbar$ is the reduced Planck constant. For the total decay rate of levels 2 and 4
we have $\Gamma = \Gamma_{12} + \Gamma_{23} =\Gamma_{14} + \Gamma_{34}$, respectively.

Employing the rotating wave approximation one obtains the following optical Bloch equations for the density matrix of the four-level system
\begin{subequations}
\label{eq:bloch}
\begin{align}
  \dot{\rho}_{11} & =  i \ \frac{\Omega_{12}}{2} \ (\rho_{21} - \rho_{12}) + \Gamma_{12} \
    \rho_{22} + \Gamma_{14} \ \rho_{44} \\
  \dot{\rho}_{22} & = -i \ \frac{\Omega_{12}}{2} \ (\rho_{21} - \rho_{12}) - \Gamma \
    \rho_{22} \\
  \dot{\rho}_{33} & =  i \ \frac{\Omega_{34}}{2} \ (\rho_{43} - \rho_{34}) + \Gamma_{23} \
    \rho_{22} + \Gamma_{34} \ \rho_{44} \\
  \dot{\rho}_{44} & = -i \ \frac{\Omega_{34}}{2} \ (\rho_{43} - \rho_{34}) - \Gamma \
    \rho_{44} \\
  \dot{\rho}_{12} & = - i \ \frac{\Omega_{12}}{2} \ (\rho_{11} - \rho_{22})
    -(\frac{\Gamma}{2} + i \ \Delta_{12}) \ \rho_{12} \\
  \dot{\rho}_{13} & = i \ \frac{\Omega_{12}}{2} \ \rho_{23} - i \ \frac{\Omega_{34}}{2} \
    \rho_{14} \\
  \dot{\rho}_{14} & = i \ \frac{\Omega_{12}}{2} \ \rho_{24} - i \ \frac{\Omega_{34}}{2}
    \ \rho_{13} - (\frac{\Gamma}{2} + i \ \Delta_{34}) \
    \rho_{14} \\
  \dot{\rho}_{23} & = i \ \frac{\Omega_{12}}{2} \ \rho_{13} - i \ \frac{\Omega_{34}}{2} \
    \rho_{24} - (\frac{\Gamma}{2} - i \ \Delta_{12}) \ \rho_{23} \\
  \dot{\rho}_{24} & = i \ \frac{\Omega_{12}}{2} \ \rho_{14} - i \ \frac{\Omega_{34}}{2} \
    \rho_{23} - (\Gamma - i \ (\Delta_{12} - \Delta_{34})) \rho_{24} \\
  \dot{\rho}_{34} & = - i \ \frac{\Omega_{34}}{2} \ (\rho_{33} - \rho_{44})
    -(\frac{\Gamma}{2} + i \ \Delta_{34}) \ \rho_{34}.
\end{align}
\end{subequations}
The diagonal elements $\rho_{ii}$ are the populations of the four levels $i=1,2,3,4$ and
$\rho_{ij}$, $i \ne j$ represent the coherences of the corresponding transitions.

At long times $t > \Gamma^{-1}$ one can adiabatically eliminate the coherences, i.e. set
$\dot{\rho}_{ij} = 0$ for $i \ne j$, to rewrite Equations (\ref{eq:bloch}) in terms of a simple set of rate equations
\begin{subequations}
  \label{eq:rateeq}
 \begin{align}
 \dot{f}_{1}(v) & = +R_{12}(v)[f_{2}(v)-f_{1}(v)]+\Gamma_{12}f_{2}(v)+\Gamma_{14}f_{4}(v),\\
 \dot{f}_{2}(v) & = -R_{12}(v)[f_{2}(v)-f_{1}(v)]-(\Gamma_{12}+\Gamma_{23})f_{2}(v),\\
 \dot{f}_{3}(v) & = +R_{34}(v)[f_{4}(v)-f_{3}(v)]+\Gamma_{23}f_{2}(v)+\Gamma_{34}f_{4}(v),\\
 \dot{f}_{4}(v) & = -R_{34}(v)[f_{4}(v)-f_{3}(v)]-(\Gamma_{14}+\Gamma_{34})f_{4}(v),
 \end{align}
 \end{subequations}
for the velocity distributions $f_i=\rho_{ii}$ of ions in a given spin state $i$.
The resulting pumping rates have a simple Lorentzian shape,
\begin{equation}
  R_{ij}(v)=\frac{\Omega_{ij}^2 / \Gamma}{1+[2\Delta_{ij}/\Gamma]^{2}}.
  \label{eq:rate}
\end{equation}
As in the experiment two counterpropagating laser beams with same detuning $\Delta_\text{pump}$
and wavenumber $k = 2 \pi / \lambda$ are used, the Doppler shift of
an ion moving with velocity $v$ in the directions of the lasers beams is accounted for
by replacing $\Delta_{12} = \Delta_\text{pump} + k \, v$ and $\Delta_{34} =
\Delta_\text{pump} - k \, v$ in Equations \eqref{eq:bloch} and \eqref{eq:rate}.
%//where we have written explicitly the dependence on ion velocity, $v$.

\begin{figure}
  \begin{centering}
    \includegraphics[clip,width=3in]{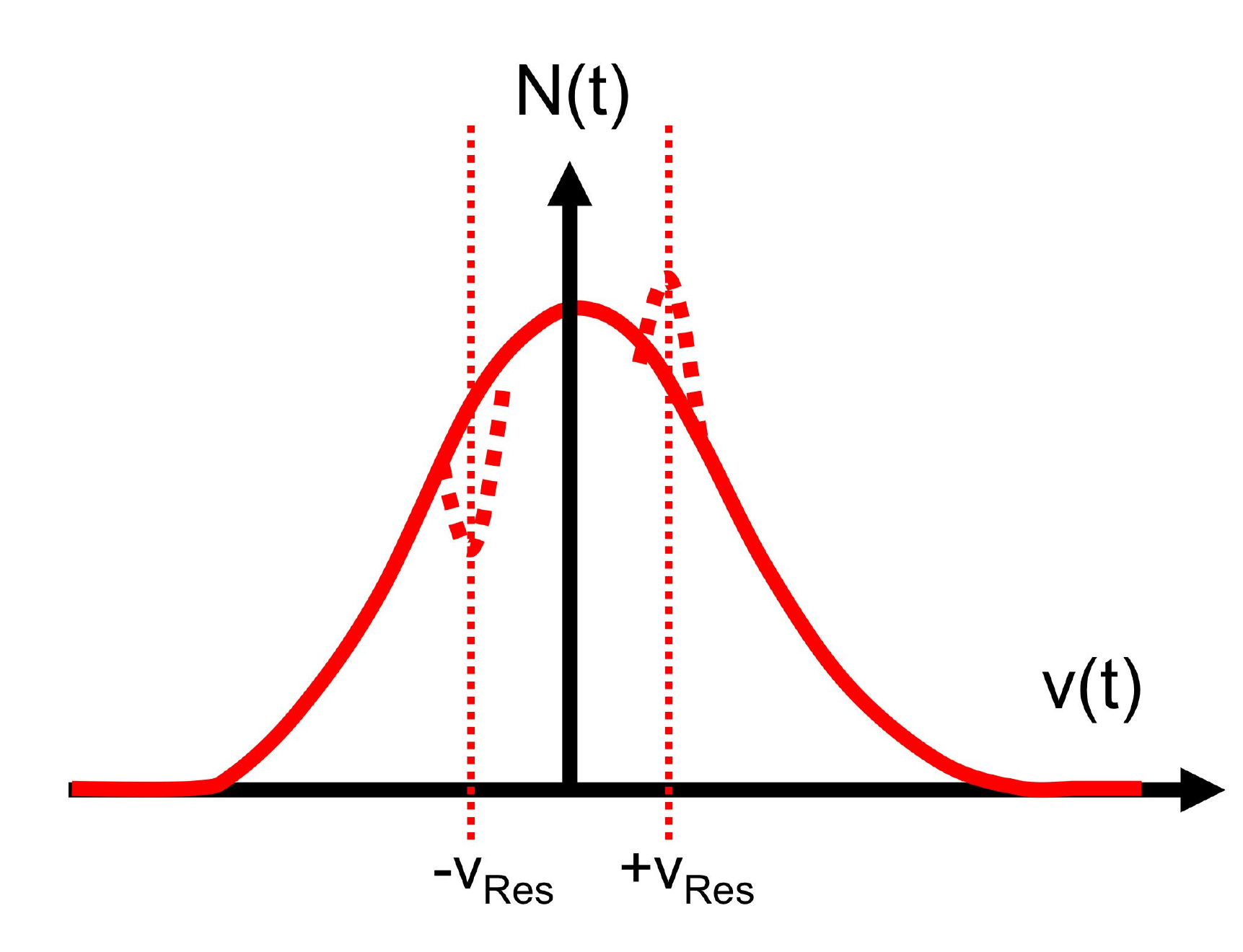}
    \par\end{centering}
  \caption{\label{fig:WholeBurning} Cartoon of the expected velocity distribution for the
    $5^2\textrm{S}_{1/2}$ $\left(m_j=+1/2\right)$, $n_3$, with both counterpropagating
    pump lasers turned on $(\sigma^-\textrm{ and }\sigma^+)$.  The solid line indicates
    the unperturbed distribution, while the dashed line indicate changes induced by
    optical pumping. A depletion of ion population is created as ions that meet the
    resonance condition $v_{Res}=\Delta_\text{pump}/k$ with the $\sigma^-$ laser while an
    enhancement in ion population appears as ions from $\left(m_j=-1/2\right)$
    $5^2\textrm{S}_{1/2}\; (n_1)$ level are optically pumped by the $\sigma^+$ laser.
    }
\end{figure}

The simplified treatment of the rate equations \eqref{eq:rateeq} shows that the laser pumping transfers spin population between
level 1 and 3 in the domain around the resonant velocities  $v_{\rm res} = \pm \Delta_\text{pump}/k$,
creating thereby an asymmetric velocity distribution as sketched in Figure \ref{fig:WholeBurning}.

Our experimental scheme to perform the described optical pumping and the resulting
velocity distributions is shown in Figure \ref{fig:exp}. The two circularly polarized pump
beams, with the same frequency yet opposite polarization and direction, interact with the
plasma for a time of $\sim100$ ns.  For the strontium ion, the
$\textrm{S}_{1/2}-\textrm{P}_{1/2}$ transition has a wavelength $\lambda = 421.7$ nm, and a total decay rate
$\Gamma / (2\pi) = 21$ MHz.  The $\sigma^{+}$ beam
has an intensity of $I_{\sigma,+}=334$\,mW/cm$^{2}$ with beam waist of $1.18$\,mm,
resulting in $\Omega_{12} / (2\pi) = 36.6$ MHz. On the other hand, the $\sigma^{-}$ pumping
beam has an intensity of $I_{\sigma,-}=315$\,mW/cm$^{2}$ with beam waist of $1.12$\,mm,
resulting in $\Omega_{34} / (2\pi) = 35.0$ MHz. Furthermore,
$\Delta_\text{pump} /(2\pi) = -20$ MHz.
For these parameters the resonant velocity $|v_{\rm res}|=8.4$ m/s is well within the
thermal velocity \mbox{$\sqrt{k_{\rm B}T/m}=15$ m/s} allowing for an efficient modification of the velocity distribution.

\section{Measurement of Perturbed Velocity Spectra}
\begin{figure}
  \begin{centering}
    \includegraphics[clip,width=3in]{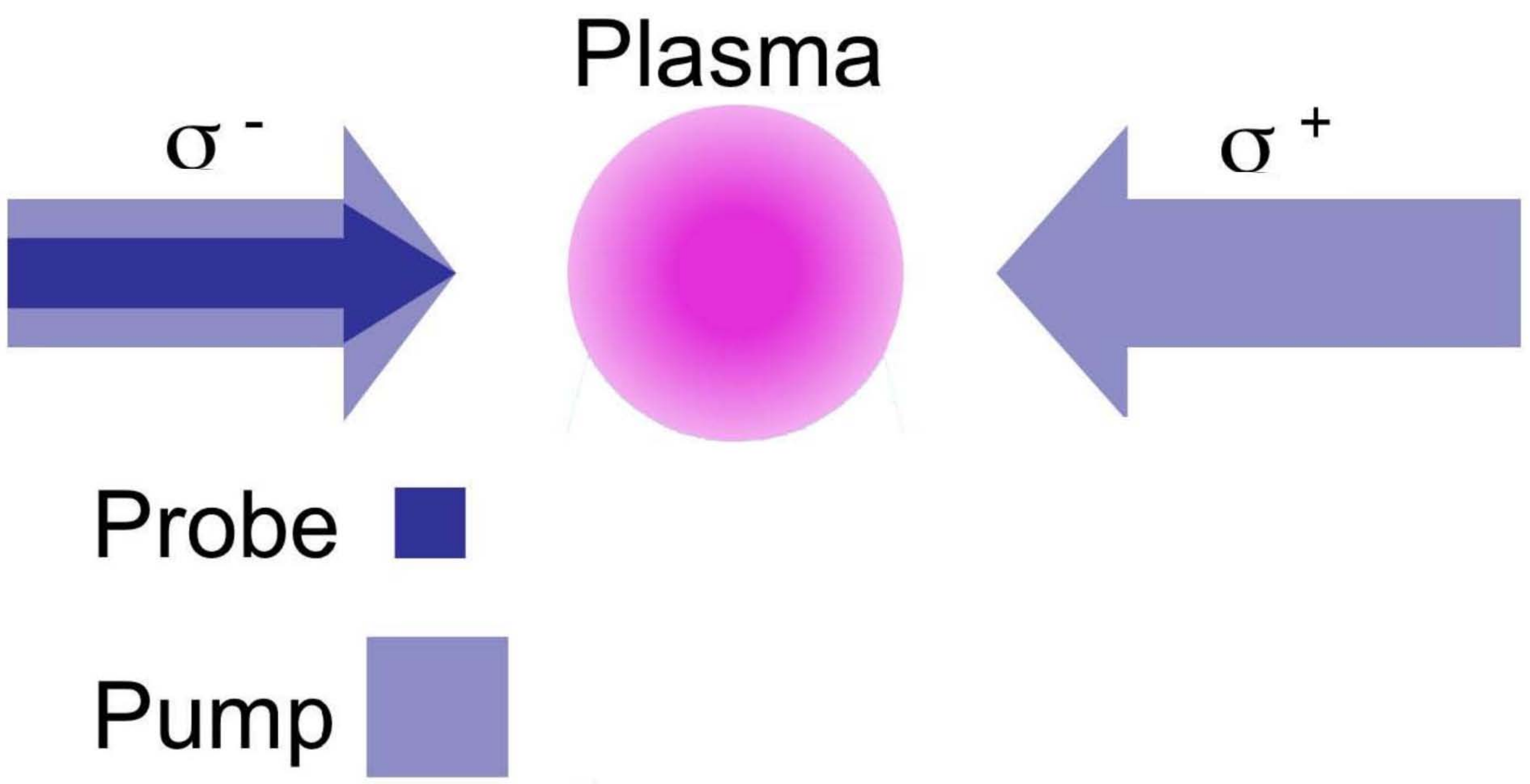}
    \par\end{centering}
  \caption{\label{fig:exp} Experimental setup for optical modification of the ion velocity distribution in an UCP. The plasma cloud is illuminated by both the pump and probe
    beams originating from a 422\,nm laser, that is near resonant with the $\textrm{S}_{1/2}-\textrm{P}_{1/2}$
    transition of \textit{Sr}$^{+}$. The pump beams have the same laser frequency yet opposite circular
    polarization and travel in opposite directions. The probe beam has
    $\sigma^{-}$ circular polarization and can be scanned to obtain a fluorescence spectrum by capturing the probe-beam induced fluorescence light with a CCD-intensified camera
    perpendicular to the beam direction.}
\end{figure}

At an adjustable time after optical pumping, a third, less intense $\sigma^{-}$ circularly
polarized probe beam, near resonant to the transition between level 3 and 4, propagates
through the plasma (see Figure. \ref{fig:exp}).
This probe beam has an intensity $I_{probe}=73$\,mW/cm$^{2}$ with beam waists of
$w_z = 0.625$\,mm and $w_x = 4.68$\,mm.
Laser-induced fluorescence in a perpendicular direction is then captured by a camera.
The signal is spatially-resolved, so we can   analyze the signal from different  regions near the center of
the plasma that have little variation in density and average plasma expansion velocity.
By scanning the frequency detuning $\Delta$ of the probe beam, we obtain a spectrum $S(\Delta)$ that contains information on the ion velocity distribution.
The measured fluorescence spectrum is proportional to the velocity-dependent population of the
ground state $f_3$ convolved with the Lorentzian excitation profile of the probe beam, i.e.
\begin{eqnarray}
  S(\Delta)&\propto&\int f_{3}(v)L(v,\Delta)dv, \label{eq:Convolution} \\
  L(v,\Delta)&\propto&\frac{1}{1+s_{0, \rm probe}+[2(\Delta-kv)/\Gamma_{\rm tot}]^{2}} \label{eq:Lorentzian}\;,
\end{eqnarray}
where, $\Gamma_{\rm tot}=\Gamma+\Gamma_{\rm inst}$ is the total linewidth,
$\Gamma_{\rm inst} / (2\pi)= 7 \,\textrm{MHz}$ is the instrumental laser linewidth and
$s_{0, \rm probe}=I_{\rm probe}/I_{\rm sat}$. The saturation intensity for the $\textrm{S}_{1/2}-\textrm{P}_{1/2}$ transition with circular polarized
light is $I_{\rm sat}=57$\,mW/cm$^2$.

\begin{figure}
  \centering
     \includegraphics[width=0.9\linewidth]{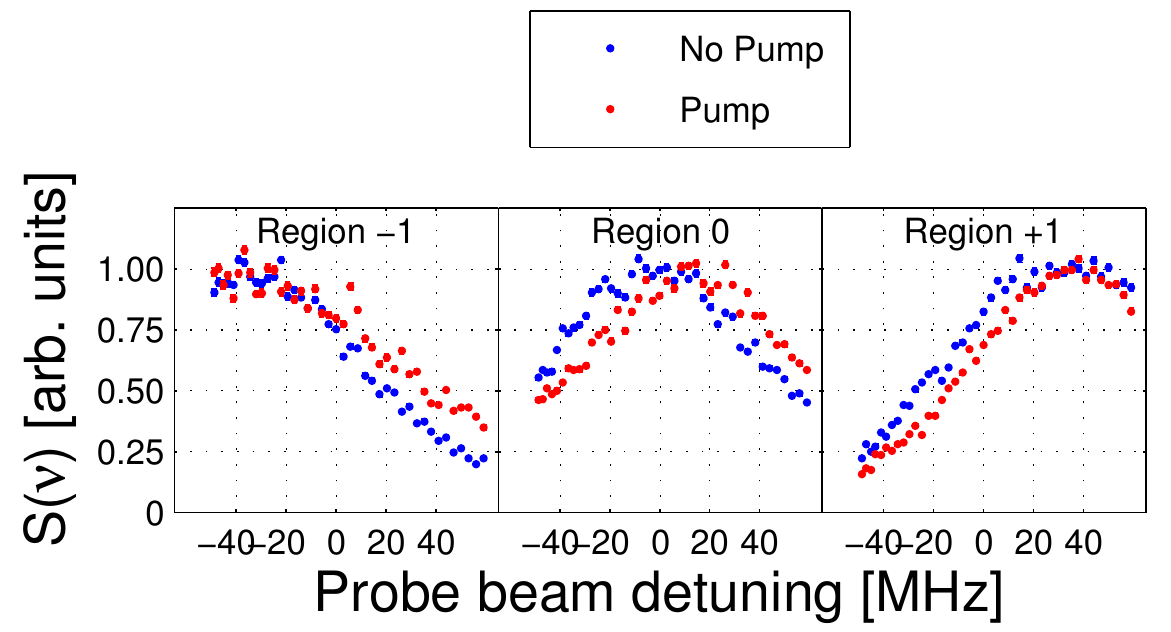}
  \caption{Ion spectra for 3 regions around plasma center with (red dots) and without
    (blue dots) pump beams for $T_i=2.3$\,K, $n_0=3.1\times10^{15}$\,m$^{-3}$ and
    $\Gamma_i=1.7$. The spectra were measured $0.13\,\mu$s after optical pumping and are
    normalized by scaling to the maximum of each curve.  To achieve low $\Gamma_i$, we
    allow the plasma cloud to expand significantly before applying the pump beams. As a
    consequence regions other than plasma center have significant expansion velocity and
    display significant shift in their spectra. Notice the depletion in ion population for
    $\Delta /(2\pi) \approx -20$ MHz and enhancement for $\Delta /(2\pi) \approx + 20$ MHz.
    \label{fig:VelDistro1st}}
\end{figure}

Figure \ref{fig:VelDistro1st} shows ion spectra measured after 100\,ns of optical pumping
and, for comparison, ion spectra without optical pumping, for a plasma with ion temperature
$T_i=2.3$\,K, density $n_0=3.1\times10^{15}$\,m$^{-3}$ and coupling parameter
$\Gamma_i=1.7$. The spectra are scaled by the maximum of each curve.
We record separately the spectra for three regions around plasma center (labeled Region -1, 0, 1).
Given the higher mean velocity in the outer regions due to plasma expansion \cite{cgk08},
their corresponding spectra are shifted from zero (Regions $\pm1$).
Notice the depletion in ion population for $\Delta /(2\pi) \approx -20$ MHz and enhancement in
ion population for $\Delta /(2\pi) \approx 20$ MHz in region 0, as discussed above.
The features here are less sharp, compared to the schematic distribution of Figure \ref{fig:WholeBurning} due
to power broadening from the pump beams \cite{mvs99} and the convolution of the velocity distribution and Lorentzian excitation profile (Equation\ \ref{eq:Convolution}).
As the probe and pump pulses originate from the same laser source, the probe pulse has a limited frequency scanning range,
resulting in the narrow range from about -50 to 50 MHz (see Figure \ref{fig:VelDistro1st}), which complicates the analysis of the experimental data.

\section{Extracting the Velocity Distribution}
As described in the previous section, the reconstruction of the underlying velocity distribution from the recorded ion spectra is
complicated due to experimental limitations on the accessible frequency range. This leads to a
loss of information contained in the wings of the spectrum.
To determine possible errors of the reconstruction process, we performed numerical simulations of the
experiment, which provide spectral information over the full frequency range, and can thus be used for accurate
 comparisons of the real and extracted velocity distribution, as obtained from a finite frequency range of the simulated spectra.

The ion dynamics including the collisional redistribution of velocities can be accurately
captured by classical molecular dynamics (MD) simulations.
The ions are represented by an one-component plasma of $N$ particles in a cubic simulation
cell with periodic boundary conditions. The mutual interactions of the ions are effectively treated by the fast multipole method
\cite{Greengard1987FMM, fmm_article, fmm}, which permits force calculations for
large particle numbers with a numerical effort that scales only linearly with the number of ions.

The optical pumping process is described by solving the optical Bloch equations (\ref{eq:bloch}) along each ion trajectory.
The internal state dynamics is included by assigning the four-level density matrix $\rho^{(q)}$ to each of the $q=1,...,N$ ions.
Initially the two ground states are equally populated, i. e. $\rho_{11}^{(q)} = \rho_{33}^{(q)} =
0.5$, and $\rho_{ij}^{(q)} = 0$ for all remaining matrix elements.
For the duration of the pump pulse we solve the optical Bloch equations \eqref{eq:bloch}
numerically for each ion individually, taking into account the time-dependent laser detuning due to the changing ion velocity obtained
from the MD simulation.
From the simulated velocity distributions $f_i(v,t)$, we obtain the fluorescence spectra by convolution with
the Lorentzian emission profile \eqref{eq:Lorentzian} according to Equation \eqref{eq:Convolution}.

With the simulated spectra at hand, we can now proceed to analyze possible schemes to
extract the velocity distribution from the finite frequency range of the experimental fluorescence spectra.
To assess the quality of the different methods, we use the average velocity $\langle v\rangle$ as a figure of merit.

\begin{figure}
  \centering
  \includegraphics[width=0.9\linewidth]{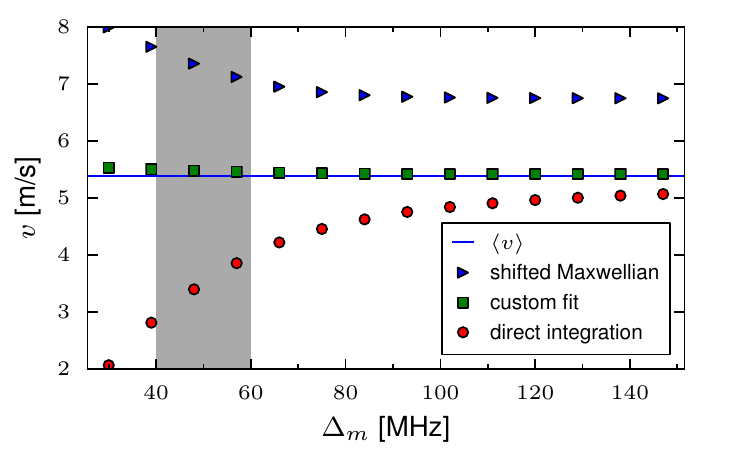}
  \caption{Comparison of the true average velocity $\vmean$ (blue line) to the average
    velocity as obtained from different extraction procedures using
    the frequency range ($-\Delta_m, \Delta_m$) of the simulated spectrum.
    Direct integration of the spectrum using Equation \eqref{eq:vd_relation} and the
    shifted Maxwellian center velocity $v_{\rm c}$ show significant deviations within the
    experimentally available frequency range  $\Delta_m / (2\pi) \approx 50$ MHz (shaded
    grey), whereas excellent agreement for the average velocity is obtained
    from the fit formula \eqref{eq:fitanalytic}.
  }
  \label{fig:vmean_compare}
\end{figure}

Given the entire excitation spectrum, $S(\Delta)$, the average velocity is related to the
average frequency shift through the wavenumber of the probe beam, $k$, according to
\begin{equation}\label{eq:exact}
  \langle \Delta \rangle = \frac{\int\limits_{-\infty}^{\infty} \Delta \, S(\Delta) d\Delta}
         {\int\limits_{-\infty}^{\infty} S(\Delta) d\Delta}= k \langle v\rangle\;.
\end{equation}
From a limited portion of the spectrum, $-\Delta_{\rm m}<\Delta<\Delta_{\rm m}$, if one
makes a Taylor expansion of the integrand in Equation \eqref{eq:Convolution} around $v=0$,
the relation can be written as
\begin{equation}\label{eq:vd_relation}
  \langle \Delta \rangle =
  \frac{\int\limits_{-\Delta_m}^{\Delta_m} \Delta \, S(\Delta) d\Delta}
    {\int\limits_{-\Delta_m}^{\Delta_m} S(\Delta) d\Delta}
  \approx \kappa(\Delta_m) \langle v \rangle
\end{equation}
The frequency-interval dependent proportionality constant is given by
\begin{equation}
  \kappa(\Delta_m) =
  k \, \left(1 - \frac{2 \, \Delta_m/\Gamma}{\left(1 + \left(2 \, \Delta_m / \Gamma
          \right)^2 \right)  \arctan\left( 2 \, \Delta_m / \Gamma \right)} \right)
    \label{eq:vd_factor}
\end{equation}
and recovers Equation (\ref{eq:exact}) in the limit $\Delta_{\rm m}\rightarrow \infty$. As shown in Figure \ref{fig:vmean_compare} this procedure yields the correct average velocity for large $\Delta_m$, but significantly underestimates the correct value of $\langle v\rangle$ for the experimentally measurable region ($\Delta_m\sim 50$\,MHz).

An alternative method, which works well for our data, is to assume a specific form for the velocity distribution and vary the parameters of the distribution to fit the resulting excitation spectrum to the experimental measurements. For an appropriate choice of the functional form of the velocity distribution, this allows  accurate approximation of the full  velocity distribution based on finite spectral information.

\begin{figure}
 \centering
 \includegraphics[width=1.0\columnwidth]{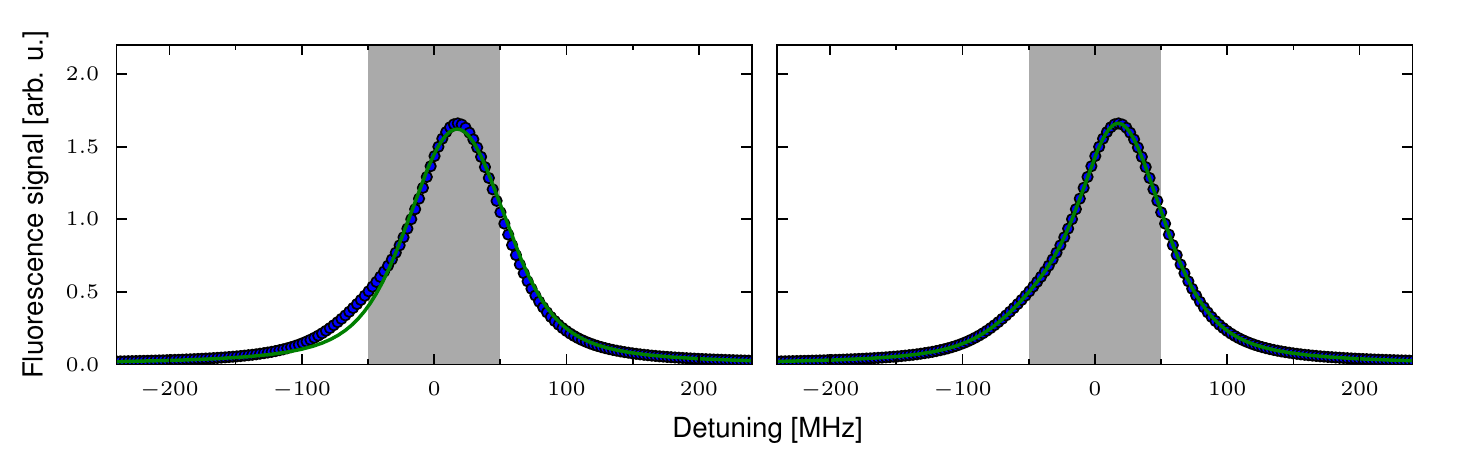}
 \caption{Fit (green line) of a convoluted Maxwellian (left) and of equation
          \eqref{eq:fitanalytic} (right) to the simulated spectrum (symbols)
          corresponding to $T = 2.8$ K, $\rho = 13 \times 10^9 {\rm cm}^{-3}$, $\Gamma = 1.4$.
          In each case the fit is only applied in the interval ($-50, 50$) MHz (shaded gray area).
          }
 \label{fig:spectrum_fit}
\end{figure}

A simple ansatz is to assume that the velocity distribution after pumping still retains its  Maxwellian shape but is only shifted by an amount $v_c$ from zero.
Figure \ref{fig:spectrum_fit}a shows the fit of a convoluted, shifted Maxwellian to the
simulated spectrum using only the experimentally available frequency range (shaded gray area).
 We find good agreement between fit and data in the central, observable part of the
 spectrum, but slight deviations at larger frequencies beyond the experimentally available
data. These small discrepancies can cause considerable deviations of the average velocity as shown Figure \ref{fig:vmean_compare}.

The situation can be improved by using a  fit formula that captures the effects of optical pumping and the collisional redistribution of velocities
more accurately. In order to construct such an expression we start from the rate equations (\ref{eq:rateeq}) and adiabatically eliminate the dynamics of the excited state populations $f_2$ and $f_4$. Collisional effects are accounted for within the relaxation time approximation \cite{bon96} which gives
\begin{subequations}
\label{eq:rates2l}
\begin{align}
  \dot{f}_{1}(v) & =  R_{\leftarrow}(v)  \ f_{3}(v) - R_{\rightarrow}(v) \ f_{1}(v) -
        \gamma \ \left[f_{1}(v) - \frac{1}{2} f_M(v) \right], \\
  \dot{f}_{3}(v) & =   R_{\rightarrow}(v) \ f_{1}(v)- R_{\leftarrow}(v)  \ f_{3}(v) -
        \gamma \ \left[f_{3}(v) - \frac{1}{2} f_M(v) \right],
\end{align}
\end{subequations}
where $\gamma$ is an effective relaxation rate and $f_M(v)=f_{1}(v) + f_{3}(v)$ denotes the Maxwellian velocity distribution at a given temperature.
Moreover, the transition rates $R_{\rightarrow}$ and $R_{\leftarrow}$ for a transition from state 1 to
state 3 and vice versa are obtained from the rate equations \eqref{eq:rateeq} according to
\begin{subequations}
\label{eq:transrates}
\begin{align}
  R_{\rightarrow}(v) & = \frac{R_{12}(v) \, \Gamma_{23}}{\Gamma + R_{12}(v)}, \\
  R_{\leftarrow }(v) & = \frac{R_{34}(v) \, \Gamma_{14}}{\Gamma + R_{34}(v)}.
\end{align}
\end{subequations}
From the steady state ($\dot{f}_{1}(v) = \dot{f}_{3}(v) = 0$) of these equations we obtain the following simple fit formula
\begin{equation}
  f_{3}(v) = f_M(v) \, \frac{ \frac{a_1}{1 + b_1 (v_{\rm res}  + v)^2} + \frac{1}{2}}
                          { \frac{a_1}{1 + b_1 (v_{\rm res}- v)^2} + \frac{a_2}{1 + b_2
(v_{\rm res} + v)^2} + 1 }.
  \label{eq:fitanalytic}
 \end{equation}

For steady state conditions, the parameters $a_1, a_2, b_1$ and $b_2$ can be expressed in
terms of the parameters of Equations \eqref{eq:rates2l} and \eqref{eq:transrates}.
In most cases, however, the system does not reach the steady state for our experimental parameters and optical pumping times. Nevertheless, one can apply Equation
\eqref{eq:fitanalytic} to fit the measured spectra using $a_1, a_2, b_1, b_2$ as free parameters. As we will show below, this ansatz indeed provides an excellent description of the actual velocity distributions.
The parameters $a_1, a_2$ quantify the importance of optical pumping relative to
collisional redistribution and the parameters $b_1, b_2$ control the widths of the asymmetric features of the velocity distribution. In the limit $a_1,a_2\gg1$, Equation \eqref{eq:fitanalytic} yields the correct steady state for very strong pumping and weak collisions. On the other hand, for $a_1, a_2 = 0$ it recovers the equilibrium Maxwell distribution, as expected in the absence of optical pumping.

As shown in Figure \ref{fig:spectrum_fit}b, the frequency spectrum obtained from
convolving Equation (\ref{eq:fitanalytic}) according to Equation (\ref{eq:Lorentzian})
yields a much improved fit as compared to the shifted Maxwellian, shown in Figure
\ref{fig:spectrum_fit}a. Moreover it accurately describes the wings of the spectrum, even
though the fitting has been determined from the finite frequency range $-50 \ {\rm
  MHz}<\Delta<50 \ {\rm MHz}$. Consequently, the extracted average velocity, shown in Figure \ref{fig:vmean_compare}, is in excellent agreement with the real value of $\langle v\rangle$ even for a very limited range of frequencies available for fitting the fluorescence spectrum. Moreover, the procedure described above can also be used to construct reliable fitting functions for more complex optical pumping schemes that may be applied in future experiments.

\section{Summary}
In this work we have presented  new techniques to perturb, probe, and model ion velocity distributions in ultracold neutral plasmas. We anticipate that the described approach will serve as a valuable tool for direct experimental measurements of non-equilibrium plasma dynamics in the regime of strong coupling.
\begin{theacknowledgments}
This work was supported by the Department of Energy and National Science Foundation (PHY-0714603).
\end{theacknowledgments}

\bibliography{literature}
\end{document}